\documentclass{article}
\usepackage{caption}
\usepackage{float}
\usepackage[T1]{fontenc}
\usepackage{graphicx}
\usepackage[utf8]{inputenc}
\usepackage{subcaption}
\usepackage[hidelinks]{hyperref}
\usepackage{xurl}
\usepackage{tikz}
\usepackage{parskip}
\usepackage{natbib}

\begin{document}
    \title{Beneficence Signaling in AI Development Dynamics}

    \author{Sarita Rosenstock
    \\
    \small{School of Computing and Information Systems}
    \\
    \small{\emph{University of Melbourne}}
    }

    \maketitle

    \begin{abstract}
        This paper motivates and develops a framework for understanding how the socio-technical systems surrounding AI development interact with social welfare. It introduces the concept of ``signaling'' from evolutionary game theory and demonstrates how it can enhance existing theory and practice surrounding the evaluation and governance of AI systems.

    \end{abstract}



\section{Introduction}
\label{sec:intro}

To conceive of something as a signal is to attend to the roles it plays in the social dynamics of information \citep{Skyrms2010-eg}. That is, rather than merely asking what a token means denotatively, we are interested in what meaning emerges from how a community uses that token for communication. This paper explains and explores what it would look like to consider the socio-technical dynamics surrounding AI development through a signaling lens. In particular, I aim to show that this perspective can help us:

\begin{itemize}
	\item identify functional roles that signals emitted in the course of AI development can/do play in shaping the trajectory of the sector towards more or less socially beneficial outcomes

	\item articulate more precisely what the goals are for ``AI policy interventions'' (which often have the form of demands or requests for particular sorts of signals), better assess how well a given strategy addresses that goal, and point towards some interesting novel strategic directions.

\end{itemize}
To start things off, I’ll first explain what I mean by ``beneficence signaling.'' I’ll then discuss why I think this is a productive avenue of research for those interested in promoting beneficial AI. I’ll proceed to sketch some of the core components of the theoretical framework I hope to build around this concept, before giving an outline of the rest of the paper. 

\subsection{Beneficence signaling}

By ``beneficence signaling'' I mean any of the myriad ways AI developers can express to one another, users, regulators, and the public that their technology and development process are conducive to socially beneficial outcomes.

Explicit signals of this sort include public statements, press releases. But a signal can be any observable output (``token'') which carries information about an individual, company, or product that can be used to inform beliefs and behaviors with respect to that entity. This includes what they say and don’t say, what they visibly do and don’t do, and how they chose to say and do these things. Examples of relevant signals include relatively explicit declarations of ethical commitment such as ``AI Ethics Principles'', agreeing to independent audits, scores on various ``fairness'' metrics, and press releases detailing harm mitigation strategies. But companies do not get to choose which of the signals they emit communicate ethically salient information— every visible output and action they make is a candidate signal for others to evaluate and use to condition their own behavior.  

Explicit statements of commitment to behave ethically do not and cannot on their own serve as signals of ethical intent. To do so, they would need to cohere robustly to actual ethical behavior by actors who send these signals in a dynamical social context. This means that to reliably send or receive a signal of something, one needs an operative model of this social context of communication, and possibly even to intervene on those dynamics.

\subsection{Motivation}

Traditional approaches to promoting safe and beneficial AI development are largely ``top-down'', centering around governance. These approaches struggle to keep up with rapidly evolving technology, making resulting legal requirements either immediately outdated or else too technologically non-specific to constitute real guarantees. The ``bottom-up'' approaches that do exist tend to focus on ways companies can unilaterally make their tech more pro-social, but do not attend to the actual incentives that might motivate companies to employ their recommendations. This paper joins a small minority of AI ethics work taking a systems-level perspective that connects and encompasses both approaches.\footnote{Connections with existing work in this field are discussed in section~\ref{subsec:GTDD}.}

The starting point of a signaling framework is the full socio-technical system--including companies, individual workers, governments, universities, think tanks, etc. Calls for transparency and accountability from governments and NGOs propagate as signals through this system and influence social and economic behavior surrounding AI, as do visible outputs from companies themselves. The fundamental insight of the signaling framework is that what a signal \textit{means} depends on the role it plays in this system in terms of sharing information and incentivizing behavior. This perspective can be useful to regulators to identify signals that are tightly associated with robust assurances of prosocial behavior by companies. 

A signaling lens also draws our attention to how these signals can serve to promote better outcomes more broadly through positive feedback loops, rather than merely looking for assessing and incentivising individual companies. For example, one can conceive of beneficial AI development as a \textit{collective action problem }wherein even well-intentioned industry actors cannot afford the costs associated with better practices \citep{Askell2019-dh}. One class of solutions to this problem can be found by analogizing work from economics and evolutionary biology regarding how signals of cooperative intent can stabilize risky cooperation and enable prosocial strategies to prevail in competitive environments \citep{Rosenstock2018-qh}.

\subsection{Features of a signaling framework}

A key feature of what makes a token an effective signal is that the token has community uptake in such a way as to actually adhere to its intended meaning. We can only verify that other tech companies are performing desired behaviors if they are sending the right kinds of signals. So, when we ask (explicitly by laws or social mandates, or implicitly by exemplary conduct) for particular behaviors, we want our ask to include or consist of sending particular sorts of signals. Exploring the features of signals we can develop and/or promote to best serve these aims is thus a promising research direction.

Before diving more deeply into the framework, I want to outline some key considerations to bear in mind throughout the report. I’m classing these under the headings Content, Uptake, Perspective, and Incentives.

\subsubsection{Content}

A core component of the theoretical contribution I hope to make here is to develop the relevant notion of ``content'' for signals of interest. In section~\ref{sec:meaning}, I attempt to assemble existing literature on signaling games into a unified account of signal content that is readily applicable to signals generated in the course of AI development that can help facilitate broadly beneficial outcomes. Much of the remaining theoretical work can be understood as articulating and responding to various challenges associated with ensuring that a signal latches on to its intended target content.

Considerations include:

\begin{itemize}
	\item Fakeability: How difficult is it to ``successfully'' send the signal without performing the associated action? 

	\item Contextuality: How does context shape meaning? How can ``the same'' signal mean different things depending on context or perspective?

	\item Dynamics: How does meaning shift over time, especially in response to perverse incentives?

\end{itemize}
\subsubsection{Uptake}

What features make signals (and associated behaviors) more likely to get community uptake in the right ways? One potential application of this framework is to consider how companies with prosocial intentions can use their positions as technological leaders to create a norm among their competitors of sending particular sorts of signals that latch onto beneficial action. For this to work, it is not sufficient for the identified or designed signals to correspond to the right material facts and behaviors; the correspondence must be robustly evolvable in the relevant fitness context (primarily capitalistic in this case). This points towards considerations of virality and information flow as not merely a second step after signal selection, but intertwined with constitutive of what a signal does and can mean. 

Considerations:

\begin{itemize}
	\item Cost: ``Cheap'' signals can be sent without substantial behavioral changes, but ``expensive'' signals might be incompatible with competitive success. 

	\item Naturalness: Should we be looking for ``natural'' byproducts of tech development to serve as signals, or should we develop novel signals designed for a particular purpose? The former is ``easier'' and requires less work to promote uptake, but greatly restricts our option space. 

	\item Simplicity: In what ways does a ``good'' signal need to be ``simple'', and how is this a problem when our goal is to signal complex states of affairs?

\end{itemize}
\subsubsection{Perspective}

A ``signaling game'' involves a \textit{sender} and a \textit{receiver}, each of whom decide what and how to communicate based on their own beliefs and motivations. While this may seem obvious, a clear consideration of who and why a signal is sent, and who is receiving it and how, is largely absent from much of the current literature in ethical AI.

Considerations:

\begin{itemize}
	\item Legibility: Signaling considerations might motivate reframing our objectives from beneficence to visible beneficence in some circumstances \citep{Benn2021-ka}.

	\item Multiplicity: Especially for public signals, how might the same token signal different things to different agents? And how can signals effectively serve multiple functions in such contexts?

	\item Receivers: Consider how signals are perceived from different standpoints, e.g.  AI workers, competitors, consumers, and regulators.

	\item Senders of interest: Consider who the relevant ``agents'' are, if any, to understand as the ``senders'' of signals, e.g. organizations, individual developers, and AI systems themselves.

\end{itemize}
\subsubsection{Incentives}

How does the inclusion of signaling in a social context influence the individual incentives of participants, and vice versa? As I will elaborate in section 3.1, a signal perspective enriches existing game-theoretic frameworks for understanding social and economic incentives from a policy perspective.

Considerations:

\begin{itemize}
	\item Cooperation: How can signals help facilitate cooperation? 

	\item Gamification: How does knowledge of the role of a signal in an incentive scheme undermine the association of signal with content, and how can we protect against this?

\end{itemize}
\subsection{Outline}

In section~\ref{sec:meaning}, I go into more detail outlining the game theoretic underpinnings of the signaling framework. In section~\ref{sec:other_work} I will draw some connections with other concepts and lines of inquiry. I conclude in section~\ref{sec:disc} by extracting some key lessons. 

\section{How signals acquire meaning}
\label{sec:meaning}

While it’s easy enough to acknowledge that the content of signals is substantially determined by context, we’re left with the difficult task of accounting for \textit{how} context determines meaning. While I doubt there is an adequate one-size-fits-all account or determination procedure we can defer to, there are plenty of formal and empirical tools we can bring to bear on particular signaling systems. I’m hopeful that these can be honed into an industry-specific theoretical toolkit that will help guide us in understanding how signals of beneficial AI acquire their meanings. In this section, I make a first attempt to assemble existing game theory literature on signals towards this end.

\subsection{The basic model}

It’s instructive to start with a simple, minimal formal model of how this can work in the form of a sender-receiver game, and slowly add complexity to show how the model can be adapted to suit different contexts.\footnote{The ensuing characterization of sender-receiver games draws heavily from Skyrms's \citeyearpar{Skyrms2010-eg} account of signaling in the context of evolutionary biology, and Franke's \citeyearpar{Franke2013-qp} linguistic framework, both building from Lewis's \citeyearpar{Lewis1969-bz} initial formulation of sender-receiver games. I deviate progressively from these existing works as I build towards an account more particularly suited to the AI development context.} We start with two agents: the \textit{sender} and the \textit{receiver}, each with their own motivations and beliefs, which we’ll examine more deeply in what follows. Keeping things simple, the sender might be in a position to observe that a coin lands heads or tails, while a receiver is not. Upon observing either $H$ or $T$, they will choose to send message $m$ or $m^*$ to the receiver. Upon receiving one or the other signal, the receiver selects an interpretation $i$ or $i^*$. 

\begin{figure}[h]
    \centering

\tikzset{every picture/.style={line width=0.75pt}} 

\begin{tikzpicture}[x=0.75pt,y=0.75pt,yscale=-1,xscale=1]

\draw (12,33.4) node [anchor=north west][inner sep=0.75pt]    {$H$};
\draw (12,78.4) node [anchor=north west][inner sep=0.75pt]    {$T$};
\draw (92,33.4) node [anchor=north west][inner sep=0.75pt]    {$m$};
\draw (92,78.4) node [anchor=north west][inner sep=0.75pt]    {$m^{*}$};
\draw (182,33.4) node [anchor=north west][inner sep=0.75pt]    {$i$};
\draw (182,78.4) node [anchor=north west][inner sep=0.75pt]    {$i^{*}$};
\draw (25,115) node  [font=\small] [align=left] {{\small sender}\\{\small observes}};
\draw (100,115) node  [font=\small] [align=left] {{\small sends}\\{\small message}};
\draw (179,115) node  [font=\small] [align=left] {{\small receiver}\\{\small interprets}};
\draw    (30,41) -- (87,41) ;
\draw [shift={(89,41)}, rotate = 180] [color={rgb, 255:red, 0; green, 0; blue, 0 }  ][line width=0.75]    (10.93,-3.29) .. controls (6.95,-1.4) and (3.31,-0.3) .. (0,0) .. controls (3.31,0.3) and (6.95,1.4) .. (10.93,3.29)   ;
\draw    (27,86.11) -- (87,86.81) ;
\draw [shift={(89,86.84)}, rotate = 180.67] [color={rgb, 255:red, 0; green, 0; blue, 0 }  ][line width=0.75]    (10.93,-3.29) .. controls (6.95,-1.4) and (3.31,-0.3) .. (0,0) .. controls (3.31,0.3) and (6.95,1.4) .. (10.93,3.29)   ;
\draw    (110,41) .. controls (111.67,39.33) and (113.33,39.33) .. (115,41) .. controls (116.67,42.67) and (118.33,42.67) .. (120,41) .. controls (121.67,39.33) and (123.33,39.33) .. (125,41) .. controls (126.67,42.67) and (128.33,42.67) .. (130,41) .. controls (131.67,39.33) and (133.33,39.33) .. (135,41) .. controls (136.67,42.67) and (138.33,42.67) .. (140,41) .. controls (141.67,39.33) and (143.33,39.33) .. (145,41) .. controls (146.67,42.67) and (148.33,42.67) .. (150,41) .. controls (151.67,39.33) and (153.33,39.33) .. (155,41) .. controls (156.67,42.67) and (158.33,42.67) .. (160,41) .. controls (161.67,39.33) and (163.33,39.33) .. (165,41) -- (169,41) -- (177,41) ;
\draw [shift={(179,41)}, rotate = 180] [color={rgb, 255:red, 0; green, 0; blue, 0 }  ][line width=0.75]    (10.93,-3.29) .. controls (6.95,-1.4) and (3.31,-0.3) .. (0,0) .. controls (3.31,0.3) and (6.95,1.4) .. (10.93,3.29)   ;
\draw    (117,87) .. controls (118.67,85.33) and (120.33,85.33) .. (122,87) .. controls (123.67,88.67) and (125.33,88.67) .. (127,87) .. controls (128.67,85.33) and (130.33,85.33) .. (132,87) .. controls (133.67,88.67) and (135.33,88.67) .. (137,87) .. controls (138.67,85.33) and (140.33,85.33) .. (142,87) .. controls (143.67,88.67) and (145.33,88.67) .. (147,87) .. controls (148.67,85.33) and (150.33,85.33) .. (152,87) .. controls (153.67,88.67) and (155.33,88.67) .. (157,87) .. controls (158.67,85.33) and (160.33,85.33) .. (162,87) .. controls (163.67,88.67) and (165.33,88.67) .. (167,87) -- (169,87) -- (177,87) ;
\draw [shift={(179,87)}, rotate = 180] [color={rgb, 255:red, 0; green, 0; blue, 0 }  ][line width=0.75]    (10.93,-3.29) .. controls (6.95,-1.4) and (3.31,-0.3) .. (0,0) .. controls (3.31,0.3) and (6.95,1.4) .. (10.93,3.29)   ;
\draw  [dash pattern={on 0.84pt off 2.51pt}]  (30,46.78) -- (87.25,78.32) ;
\draw [shift={(89,79.29)}, rotate = 208.85] [color={rgb, 255:red, 0; green, 0; blue, 0 }  ][line width=0.75]    (10.93,-3.29) .. controls (6.95,-1.4) and (3.31,-0.3) .. (0,0) .. controls (3.31,0.3) and (6.95,1.4) .. (10.93,3.29)   ;
\draw  [dash pattern={on 0.84pt off 2.51pt}]  (27,81.03) -- (87.25,47.76) ;
\draw [shift={(89,46.8)}, rotate = 151.09] [color={rgb, 255:red, 0; green, 0; blue, 0 }  ][line width=0.75]    (10.93,-3.29) .. controls (6.95,-1.4) and (3.31,-0.3) .. (0,0) .. controls (3.31,0.3) and (6.95,1.4) .. (10.93,3.29)   ;
\draw  [dash pattern={on 0.75pt off 0.75pt}]  (110,46.43) .. controls (112.25,45.71) and (113.73,46.47) .. (114.44,48.72) .. controls (115.15,50.97) and (116.63,51.73) .. (118.88,51.02) .. controls (121.13,50.3) and (122.61,51.06) .. (123.33,53.31) .. controls (124.04,55.56) and (125.52,56.32) .. (127.77,55.61) .. controls (130.02,54.9) and (131.5,55.66) .. (132.21,57.91) .. controls (132.92,60.16) and (134.4,60.92) .. (136.65,60.2) .. controls (138.9,59.49) and (140.38,60.25) .. (141.09,62.5) .. controls (141.8,64.75) and (143.28,65.51) .. (145.53,64.79) .. controls (147.78,64.08) and (149.26,64.84) .. (149.98,67.09) .. controls (150.69,69.34) and (152.17,70.1) .. (154.42,69.38) .. controls (156.67,68.67) and (158.15,69.43) .. (158.86,71.68) .. controls (159.57,73.93) and (161.05,74.69) .. (163.3,73.98) .. controls (165.55,73.26) and (167.03,74.02) .. (167.74,76.27) -- (170.12,77.5) -- (177.22,81.17) ;
\draw [shift={(179,82.09)}, rotate = 207.33] [color={rgb, 255:red, 0; green, 0; blue, 0 }  ][line width=0.75]    (10.93,-3.29) .. controls (6.95,-1.4) and (3.31,-0.3) .. (0,0) .. controls (3.31,0.3) and (6.95,1.4) .. (10.93,3.29)   ;
\draw  [dash pattern={on 0.75pt off 0.75pt}]  (117,79.42) .. controls (117.67,77.16) and (119.14,76.37) .. (121.4,77.04) .. controls (123.66,77.71) and (125.12,76.92) .. (125.79,74.66) .. controls (126.46,72.4) and (127.93,71.61) .. (130.19,72.28) .. controls (132.45,72.95) and (133.92,72.16) .. (134.59,69.9) .. controls (135.26,67.64) and (136.73,66.85) .. (138.99,67.52) .. controls (141.24,68.19) and (142.71,67.4) .. (143.38,65.15) .. controls (144.05,62.89) and (145.52,62.1) .. (147.78,62.77) .. controls (150.04,63.44) and (151.51,62.65) .. (152.18,60.39) .. controls (152.85,58.13) and (154.32,57.34) .. (156.58,58.01) .. controls (158.84,58.68) and (160.3,57.89) .. (160.97,55.63) .. controls (161.64,53.37) and (163.11,52.58) .. (165.37,53.25) .. controls (167.63,53.92) and (169.1,53.13) .. (169.77,50.87) -- (170.21,50.63) -- (177.24,46.82) ;
\draw [shift={(179,45.87)}, rotate = 151.58] [color={rgb, 255:red, 0; green, 0; blue, 0 }  ][line width=0.75]    (10.93,-3.29) .. controls (6.95,-1.4) and (3.31,-0.3) .. (0,0) .. controls (3.31,0.3) and (6.95,1.4) .. (10.93,3.29)   ;

\end{tikzpicture}

    \caption{\small{Illustration of possible solutions to sender-receiver game. Sender has two options to encode message, shown by solid and dotted lines. Receiver has options for decoding message, shown by solid and dotted wavy lines.}}
    \label{fig1}
\end{figure}
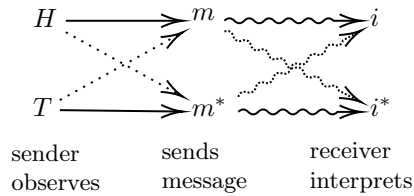

As shown in figure~\ref{fig1}, the sender decides which signal to send by considering how they expect the receiver to respond– i.e. based on their model of how $m$ and $m^*$ correspond to $i$ and $i^*$ for the receiver, and whether the sender hopes to elicit $i$ or $i^*$. Which interpretation--$i$ or $i^*$--is preferable to the receiver depends on whether the coin landed heads or tails, which they can only infer indirectly via their model of how the sender associates $H$ or $T$ with $m$ or $m^*$.

Let’s start with the easy case where the sender just wants the receiver to know the truth about whether the coin landed heads or tails. If both agents share a language and trust in one another’s honesty, it’s easy to see what they should do. Upon observing either Heads or Tails, sender will send the message $m =$ ``The coin landed Heads'' or $m^*=$ ``The coin landed Tails'' respectively.  The receiver will interpret the message as meaning $i =$ The coin \textit{actually }landed Heads or $i^* =$ The coin \textit{actually} landed Tails.

If the sender and receiver don’t share a language, the sender and receiver will still have to decide on strategies for sending and interpreting messages. A strategy for the sender will be an association between stimuli and messages. In the absence of any clues as to how the receiver will interpret their messages, the sender will be indifferent between the association $H \mapsto m$ and $T \mapsto m^*$, and the association $H \mapsto m^*$ and $T \mapsto m$. The receiver will be similarly indifferent between the interpretations [$m \mapsto H$ \& $m^* \mapsto T$] and [$m \mapsto T$ \& $m^* \mapsto H$]. This gives us two senses of the ``meanings'' of the signals $m$ and $m^*$: one from the sender’s encoding, and one from the receiver’s interpretation. 

\subsection{From model to meaning}

Both sender and receiver require a \textit{strategy} for forming this sort of association between states and messages. Since we’re starting with the simplifying assumption that the sender and receiver are motivated by the shared goal of conveying accurate information, as long as one or the other is able to access some feedback regarding how well communication is working, they will eventually settle on a shared language [$m \leftrightarrow H$ \& $m^* \leftrightarrow T$] or [$m \leftrightarrow T$ \& $m^* \leftrightarrow H$]. Such feedback may take the form of deliberate strategic adaptation to new information, or selection effects associated with aligned communicators out-competing misaligned ones.\footnote{A lot of the language around this, e.g. ``selection'', ``evolution'', ``fitness'', betrays the origins of evolutionary game theory in ecology and biology. For our purposes this is somewhat metaphorical, with the selection pressures being largely economic/cultural, though there are debates about the aptness of this analogy \citep{Lewens2015-gm}.} 

In a corporate context like this one, it’s tempting to speak only using the former, ``strategic'' terms. The relevant actors, after all, involve humans (or groups of them) with the rational faculties required to calculate the optimal course of action for themselves. Indeed, rational best response is a powerful lens, and there are particular decision-points that are best understood in these terms, and identifying these points should be a core goal of this endeavor. But this is a mistake for a number of reasons. As we stray further from the basic model, the notion of a ``rational best response'' loses sense as the calculations involve more interacting parameters with high variance, and the only sensible thing a truly rational actor can do is attend to more holistic features of the evolutionary dynamics in which that singular interaction takes place. Moreover, conceiving of signaling strategies as purely strategic can lead to inappropriately associating signal content with the intention of the sender. If signals are intentional, then ``misleading'' signals are adversarial, and thus to question meaning is to accuse a sender of bad intentions. But signals can deviate from sender intention for lots of reasons--a sender might not even \textit{have} an intention associated with a signal--and we should be wary of falling into this trap.

It therefore behooves us to consider both mechanisms–individual strategic behavior and evolutionary selection–in order to understand how an initial communication ``problem'' resolves into a ``solution'' in the form of a stable language associating signals with meanings. To see how these work, we begin by defining the ``basic model'' as a game in which both sender and receiver get a constant ``reward'' whenever the receiver is able to correctly ascertain whether the coin landed heads or tails. 

If both actors have sufficient reasoning capabilities, they will almost certainly settle on a correct language more or less quickly. If the first message ($m$) by pure chance leads to the correct association ($H$), then both sender and receiver will be motivated by their mutual reward to continue with this association in future interactions. If the association fails at first, they will try again until something sticks. Note that there is a potential for a vicious cycle here, whereby both parties ``swap'' associations in light of failures and never find success. In this simple game, the cycle can be dislodged easily by any amount of experimentation in strategy by the players in signal selection or interpretation, unless they are extraordinarily unlucky and all such attempts are hopelessly synchronized. But it is worth attending to nonetheless, as such accidental failures to communicate by rational agents with aligned intentions often are thwarted by chance, though this chance may be minimized by agents’ persistent strategic experimentation.

By contrast, an evolutionary perspective does not require any reasoning or strategy on the part of the players. Rather, we merely interpret the ``reward'' in terms of ``fitness'', whereby agents with higher rewards are somehow more likely to ``persist'' in the ecosystem. The use of game theory in this way was initially used to explain biological evolution, but evolutionary game theory turns out to be useful to economists and social scientists as well, for understanding which companies ``survive'' on the market, or which cultural practices ``survive'' in a community. There are a number of ways we can formalize this idea mathematically in terms of \textit{dynamical evolution} over time. 

A simple, standard way to do this is the \textit{replicator dynamics}, by which successive generations of actors \textit{replicate} the strategies of previous generations in accordance with how well rewarded they were. We can imagine the first generation selecting an association between message and state purely at random. Some will succeed and some will fail, and this will be reflected in greater and lesser numbers of each strategy in the following generation. It is technically possible for both languages to evolve simultaneously in this manner--especially if the dynamics is \textit{localized} and agents are more likely to interact with their neighbors. But generally over time, the replicator dynamics will amplify slight advantages that might appear from one language being slightly more common, until the community settles on a single language.

Note that the presence of background evolutionary dynamics is perfectly compatible with a rational-strategic understanding of agent behavior. We might interpret the dynamics as at least partially deriving from intentional strategy selection, or we might imagine a background dynamical system within which one might gain advantage by behaving more strategically. These sorts of intermediate perspectives seem most appropriate for the case at hand. 

\subsection{Correlation mechanisms}

For the basic model, the resulting language will be purely \textit{conventional} in that neither association between messages and contents is more likely or apt to solve the problem of aligned meanings. The symmetry of the set-up that led to this pure conventionality is an idealization that will typically not hold--it’s more accurate to view conventionality as variable  \citep{OConnor2021-bb}. The symmetry might be broken in a number of ways. This is most obvious if we consider that the agents might have \textit{some} information about their interlocutor that will cause them to expect one association to be better aligned. 

We can also expect asymmetries as a result of including two additional parameters in our model: the \textit{costs} associated with each message $c(m)$ and $c(m^*)$, and the sender and receiver’s respective \textit{prior beliefs} regarding the bias of the coin, represented as probabilities $P_{S}(H)$ and $P_{R}(H)$. If there is any difference in cost between messages, and any difference in priors, we would expect utility maximizing agents to ``prefer'' (again, either in the strategically deliberate or passive evolutionary sense) the language that associates the costlier signal with the less likely outcome.  

These sorts of symmetry-breaking features of a game-theoretic interaction can be understood as \textit{correlation mechanisms} \citep{Hamilton1964-mz}, \textit{focal points} \citep{Schelling1960-ky}, or invocations of \textit{salience} \citep{Lewis1969-bz}. Though there may be multiple choices of actions or linguistic frameworks that my co-player can choose from, if we have sufficient understanding of one another’s motivations, we can take advantage of shared observations to independently arrive at a compatible, optimal solution.\footnote{See \citet{Aumann1987-ki} for a formalization and proof.} The correlation mechanism discussed above refers to exogenous environmental features--the costs and frequencies in our shared environment. Shared social norms and laws can also function as correlation mechanisms. And by using \textit{pre-play signaling}, we can arrive at more fine-grained mutual assurances behavior, thereby making it easier to productively cooperate \citep{Santos2011-he}. 

\subsection{Ambiguity and uncertainty}

Consider also how incorporating \textit{signal fidelity} can impact our analysis. There might be some noise in the observation and/or communication channels, so that trust and shared language are not fully sufficient for the receiver to be confident in their interpretations. In such cases, it might be worth adding an additional ``layer'' to our model to separate \textit{message interpretation} (evaluation of the sender’s intention behind sending the signal) and either \textit{belief formation} on the basis of their interpretation, and/or \textit{reaction} (action performed on the basis of interpretation, which may or may not be relevantly mediated by a belief formation). 

Further, note that shared language will require a shared understanding of possibilities for states and messages, which is a non-trivial assumption. Since signal content is a holistic product of solutions to these sorts of signaling games, content has just as much to do with which messages \textit{were not} sent as the ones that were. Even cooperative communication partners might struggle to contend with the additional coordination problem associated with aligning expectations in this way. It is interesting to consider whether and how communication can occur with only partial model alignment here. Further, if there are more states than signals to choose from (e.g., because there are meaningfully different ways an AI system might satisfy some quantitative fairness criterion), any resulting language will have to associate multiple meanings with a single signal.

These considerations are relevant for us even in cases of pure communicative cooperation, but they loom larger when the goals of communication can diverge and conflict. Deceitful communication partners can weaponize ambiguity in many ways.

First, deceitful actors can leverage and overstate uncertainty about the nature and degree of harm from their actions in order to evade criticism and attempts at regulation. This sort of ``manufactured doubt'' has been well-documented in the history of tobacco and oil and gas industries \citep{Oreskes2011-nl}. These cases can be interpreted as effective propaganda campaigns to re-interpret the ``signals'' generated by scientific studies as less indicative of the presence of problems and the nature of solutions. Even without antisocial actors or intentions, expressions of uncertainty from experts regarding the evidentiary basis for policy proposals can erode the potential for such evidence to serve as a strong signal upon which we coordinate effective public actions  \citep{Ojea_Quintana2021-dd}.

Malicious ``doubt-mongering'' can be understood as a particular instance of a more general phenomenon of abusing norms of discourse. The concept of ``conversational implicature'' originates from  \citet{Grice1975-bd}, who argued that communication often relies on the \textit{cooperative principle}: ``make your conversational contribution such as is required, at the stage at which it occurs, by the accepted purpose or direction of the talk exchange in which you are engaged.'' The point is that the content of what we say often cannot be taken at face value from \textit{what} we say, but by considering \textit{how} we have chosen to say it. Doubt-mongering can be understood as abusing Grice’s ``maxim of quantity'', whereby we generally presume the amount of information we are given is appropriate to a shared goal of discourse. When reasons for doubt are emphasized and expounded upon, while reasons for belief are barely mentioned, we assume it is because this ratio appropriately captures the speaker’s understanding of available evidence, and is the appropriate balance for decision making. Doubt-mongers can thus \textit{imply} that there is more reason to doubt than there is without ever explicitly saying anything untrue. 

Tech companies can similarly abuse implicature norms to generate signals that are misleading but not untrue. For example, they might emphasize their product’s positive performance on some metric while conspicuously omitting information about its poor performance on relevant and related metrics. This is in fact so accepted a practice from corporations that it is not usually classified as deceptive, merely ``business as usual''. However, companies regularly weave back and forth between attempting to invoke a cooperative ``conversational'' context and retreating to a more adversarial ``self-interested business'' context when convenient. It might thus behoove us to consider information manipulation theory \citep{McCornack1992-dh} to analyze how companies can exploit presumed context to manipulate the perceived content of the signals they emit.

\subsection{Complicating motives}

As is becoming clear, the story gets significantly more complicated once we allow that our agents might have other motivations beyond effective communication. Their motives need not be adversarial for this to be an issue. For example, once we add in the costs associated with sending signals, a sender’s desire to send accurate information might be overshadowed by their motivation to avoid the costs associated with sending particular signals. Incorporating other incentives beyond perfect communication, which are arguably always present, changes how signals can acquire meanings in the first place, and can also shift the pragmatic content of established languages. 

The concept of a ``misleading'' or ``deceptive'' signal presents a \textit{prima facie} philosophical puzzle. Insofar as the ``meaning'' of a signal is entirely determined by the conditions that lead to the signal being sent, there is a sense in which signals are inherently honest  \citep[p.~74]{Skyrms2010-eg}. For deception to occur, it must leverage a background context of expected usage, one that can be eroded by repeated exploitation. Evolutionary dynamics and rational learning thus afford us with a sort of immunity to sustained, large-scale deception. It is especially instructive for our purposes to consider some of the ways in which deceptive signaling can nonetheless persist.

First, a default expectation of honest signaling and cooperative behavior can be adaptive despite the presence of some antisocial behavior, so long as it is contained in a way that prevents it from spreading and taking over a population.\footnote{Formally, this can be characterized as a pseudo-separating equilibrium  \citep[p.~1003]{Huttegger2015-wh}} Thus even though deceived agents may experience one-off punishments for ``trusting'' a signal, this need not fully undermine the signal’s meaning in general. For example, if one unscrupulous vendor tells me they are selling me 5 kg of rice that turns out to in fact weigh 2 kg, I will not thereby conclude that the label ``5 kg'' really only \textit{means} less than half that amount, but confine the broken association of the term to the vendor who deceived me. So long as such experiences are sufficiently rare, I can continue to presume that grocery labels will have a particular factual relationship with their contents. But too many egregiously deceptive labels will make it unwise for me to expect any association between label and contents, thus undermining all signaling functions of labels.  

Unscrupulous vendors are thus more likely to get away with small perturbations of usage. This is easier to accomplish with vaguer terms than metric units, like ``unit'' or ``bushel.'' When there is tolerance for these sorts of unobtrusive changes in usage, the meaning of terms \textit{will} dynamically evolve alongside (subtly) deceptive usage, as with the phenomenon of shrinkflation. This indicates how a kind of \textit{dynamical deception }can persist despite the aforementioned ``immunity'' that systems naturally have to the widespread deceptive use of signals with static meaning. In an arms-race fashion, by the time I have adjusted to your deceptive use of a signal, you might respond to my adjustment with further deception. So long as it is (perceived as) sufficiently useful for me to maintain some degree of association between signal and state, perpetual deceptive use can continue without fully breaking the connection between signal and meaning. 

But what counts as a sufficiently ``minor'' deception to ``pass'' through the filter of deception-resistance can vary based on other factors. As mentioned, it can be rational for me to leave myself vulnerable to major deceptions so long as I believe they are sufficiently unlikely that I still expect to benefit from presuming honesty. Deception can thus persist if my information environment is compromised, and my assessment of the likelihood of deception is miscalibrated. 

This mechanism of deception–resulting from asymmetric information access, and the possibility of manipulating others’ access to information–will be especially important for us to attend to here. AI developers will almost universally have more access to information than users or other stakeholders attempting to interpret their signals, and the particular nature of AI capability threatens to widen this gap in unprecedented ways.

A further explanation for the persistence of deceptive signaling is the fact that signals often appear to us in contexts involving multiple game theoretic interactions. A deceptive salesman mis-applying a weight label will not change my understanding of the use of metric units because in my broader social context, metric units are widely used in straight-forward contexts in which the ``basic model'' is reasonable to assume—both parties often \textit{are} motivated primarily to cooperatively share information. I can maintain a healthy skepticism and not take for granted that this the basic model always applies, but nonetheless misidentify which circumstance a given interaction falls into. In this way, AI developers might  cleverly (or even accidentally) ``launder'' signals through academic or government communication channels, thus attaching a veneer of cooperative epistemic motive to what is in fact a potentially adversarial profit motive. This is a particular problem for AI because the cutting edge of knowledge accumulates in corporate settings, creating a knowledge asymmetry that can foster epistemic dependence. 

In sum, even \textit{with} robust self-correcting dynamics making deceptive signaling near oxymoronic, individual agents can persistently deceive by leveraging perpetual knowledge advantage of the contexts and consequences of certain signals, intentionally or otherwise.

\subsection{Cost, fakeability, and commitment devices}

The forgoing discussion makes the phenomenon of ``deceptive signaling'' sound more nefarious and deliberate than it necessarily is. It is important to keep in mind that it is quite natural for agents to respond primarily to their fitness contexts, rather than always deferring to some higher calling to ``communicate truthfully''. This is important to keep in mind when considering deliberate beneficence signaling from AI developers, such as adopting ethical principles. These signals can be understood in the context of multiple different signaling games. If read as statements of factual intention on the part of the particular people asserting them, they may in fact be perfectly ``honest'' acts of communication. This is consistent with them being completely useless as signals from the \textit{company} about expected corporate behavior. If the right sorts of dynamical constraints are not in place within companies and the economic and social contexts in which they act, the relationship between the signals intended by the authors and those in fact sent by the issuing of the document are likely not perfectly identical. 

Is there a way to distinguish between two agents asserting the same ethical commitments, one who is honestly reporting their intended behaviors, and another one who is merely saying what needs to be said to be well-received by customers and regulators? The short answer is we can’t; ``intentions'' are not the sorts of things that can be detected. 

One thing we can do is look for, engineer, encourage, and rally around signals that are ``unfakeable'' where possible—signals for which the very act of (successfully, compellingly) sending them is difficult or impossible to send if the relevant state of affairs does not hold. Fakeability generally exists on a continuum, with some signals being more or less intrinsically tied to content. In a biological context, Thomson gazelle’s ``stotting,'' or jumping high when they see a predator, is often cited as a relatively unfakeable signal of being agile and difficult to chase. In contrast, poisonous frogs use bright coloration to signal undesirability to predators, a signal that can be and is ``faked'' by non-poisonous frogs who gain the signaling benefit (while also decreasing the effectiveness of the signal in terms of predator expectations). 

Where signals are theoretically fakeable, having it ``cost'' agents to send a signal can fill the gap to help them be viable nonetheless, so long as the cost is not too burdensome to honest signalers. \citet{Huttegger2015-wh} rightly point out that the distinction between ``cost'' and ``fakeability'' is not clear cut; gazelle stotting \textit{is} costly from a fitness perspective, it’s just a cost that is more naturally understood as already ``budgeted for'' in predator evasion, which requires bursts of speed and agility regardless, whereas coloration feels like a more ``artificial'' cost imposition. Similarly in an economic context, we tend to think of taxes and subsidies as costs, as opposed to more diffuse impacts of reputation and relationship enhancing actions, despite these forces’ ability to have the same ultimate impact in terms of costs/profits of doing business. 

The moral here is that ``costliness'' and ``unfakeability'' are better understood as two different lenses which can be more or less intuitive to use, but serve the same function for behavioral incentives. \citet{Rosenstock2018-qh} demonstrate this with a mathematical model, showing that in an iterated prisoner’s dilemma game in which players can send an ``apology'' signal upon defection to indicate intent to cooperate in the future, cost and fakeability can be used interchangeably to allow cooperative behavior to evolve despite the threat of deceptive apologizers. Importantly, those results also rely on sufficient repeat interactions with the same actors, so that signals can usefully influence strategy. They also rely on a kind of ``cultural starting point'' in which actors are willing to entertain apologies despite the risk of exploitation—cooperative behavior cannot evolve from nowhere without some people sticking their necks out a bit, so it can be difficult to bootstrap. 

\subsection{Evolving meaning}

It’s vital to keep in mind that the meaning of signals is never fixed. The system is always evolving, and the very act of signaling in a dynamical evolutionary context changes what a signal ``means'' by changing the associations other actors make with it. Applying a signaling framework to identify a signal’s contextual meaning therefore requires humility and finesse. It is not enough to determine that a signal functionally associates with a particular ``ground truth'' in one context, but we must recognize the context sensitivity of that association, and take care to ensure the relevant conditions are met in other cases for the association to hold. 

Meaning also ``evolves'' in predictable, characteristic ways which we should anticipate. One common pattern is that embodied in ``Goodhart’s Law,'' an adage often stated as ``when a measure becomes a target, it ceases to be a good measure.'' The concept of a measurement here functions as a publicly recognized signal. The idea is that to initially be identified as a ``measurement,'' it would have once been a strong signal. But as soon as the signal is rewarded in the system in any way that is theoretically separable from the content, the signal’s meaning will be eroded by reward-seeking. A notorious example is the cobra effect \citep{Siebert2001-mo}, in which a British Colonial policy of providing bounties for dead cobras, aimed at reducing the cobra population, perversely encouraged cobra breeding. While not always backfiring so dramatically, leaning too heavily on a particular observable indicator to inform policy choices is always subject to some vulnerability along these lines due to the shifting meaning of signals.

Over-reliance on single evaluation metrics is a recognised problem in AI  \citep{Raji2021-lv,Liao2021-yc}. A signaling lens may explain how and why these failures occur, and help direct us towards solutions. While evidence from dynamics models can be sensitive to modeling choices and difficult to interpret \citep{Rosenstock2017-ub}, one of the most robust findings in this domain is that communities are more effective at uncovering the truth if they employ a diversity of methods and entertain a variety of hypotheses, rather than rapidly agreeing on a uniform set of methods and beliefs \citep{Zollman2010-wy}. Another direction suggested by this literature is to look for signals that emerge as best responses to general classes of reward games rather than specific instances, which can be more easily gamified or even non-maliciously over-fit  \citep{Skyrms2010-wy,Zollman2008-re,Mengel2012-eg}.

\section{Relation to other work}
\label{sec:other_work}

\subsection{Game theoretic development dynamics}
\label{subsec:GTDD}

The most obvious academic cohort which can benefit from incorporating signaling considerations is the large body of work on game theory informed policy analysis \citep{Hermans2014-qr}, especially for economic policy \citep{Phlips1995-vp} and policy surrounding the development of dangerous technologies \citep{Coe2020-xy}. There is some more recent work using game theory to analyze AI development policy in particular.

\citet{Askell2019-dh} argue that if developing AI responsibly incurs any extra cost above doing so irresponsibly, AI developers who aim for beneficial outcomes are faced with a prisoner’s dilemma. The article suggests high-level features of policy interventions that show promise for addressing game-theoretically similar situations. As incorporating signals can make iterated prisoner’s dilemmas more tractable \citep{Rosenstock2018-qh,Santos2011-he}, these would be a natural and advantageous addition to the base theoretical model. 

In a series of ongoing papers, Robert Trager and colleagues develop a more precise game-theoretic framework for modeling the same phenomenon, which they call ``dangerous technology races,'' which they use to illustrate how various features of the AI development context can contribute to the achievability of positive outcomes. One of the mechanisms explored is ``knowledge sharing,'' which is modeled as a raw ``amount'' of information about the technology that is shared among AI developers  \citep{Stafford2022-ub}. One way to incorporate signaling here is to adapt this modeling feature from a categorical or one-dimensional variable to a set of more complex options about the nature and method of knowledge sharing. After all, knowledge is not a commodity to be straightforwardly ``shared'', but is constituted by a body of evidence along with a framework for incorporating that evidence into speculative inferences about how technology works. Different ``slices'' of that conglomeration will communicate different ``facts,'' and it behooves us to recognize this in our models.

One can also understand signals as appearing here and in  \citep{Han_The_Anh2022-bm} in the form of ``commitments'' to cooperative behavior. So signals are not entirely absent, but as I discuss in the next section, this may not be the most fruitful way to incorporate them in our models.

\subsection{Credible commitments}

Insofar as ``signaling'' features in existing literature on this topic, it is usually narrowly scoped to consider signals of a particular sort: stated commitments to perform a certain action or abide by a set of rules. To view these \textit{as} signals, we are forced to contend more explicitly with certain contextual features of the situation. 

First, in drawing out the sender and receiver roles, we need to ask ``Who are \textit{we} to demand particular sorts of signals, and who are \textit{you} to supply them?'' What if ``we'' are not in an especially powerful position to make demands on AI organizations? Thinking about signals more broadly expands the scope to include things like ``natural'' signals that emanate from organizations just going about their business, as well as other sorts of ``artificial'' ones (e.g., agreeing to use model cards or allow auditing) that latch onto and encourage the sorts of behaviors we would otherwise ask for ``commitments'' to. 

Even if we do manage to extract commitments, assessing ``credibility'' might be difficult in an AI context, where tasks are often quite complex and difficult to evaluate. Asking how to make the \textit{commitments} credible may not be the right question, and it could be useful to instead ask how to assess trustworthiness outside of the context of explicit commitment-making. The ``credible commitment'' framing puts us in a position where of course everyone wants to get the benefits associated with \textit{making} the commitment without incurring the costs of following through, and it frames the problem as one of how to set up the reward structure so that’s difficult to pull off.

\subsection{Trustworthiness}

In light of the above considerations, rather than looking to evaluate the credibility of particular claims, we might look for higher level indications of \textit{trustworthiness} of other actors more broadly, over and above mere adherence to some particular set of rules and guidelines. This is supported by work referenced earlier suggesting that cooperation is more stable when we use general strategies for classes of games rather than ``fine-tuned'' strategies. So what could a general and effective ``trustworthiness'' signal look like? 

Hardin \citeyearpar{Hardin2002-oa} characterizes trustworthiness in terms of ``encapsulated self-interest''—I can trust someone if it is in their self-interest to take my self-interest into account. Regardless of the adequacy of this account, it is not readily conducive to a signaling framing. To assess you as trustworthy, I need to determine what motivates you, and (as discussed below in the context of intent) this is not directly observable. Jones’ \citeyear{Jones2012-aw} account is more promising here. 

``B is trustworthy with respect to A in domain of interaction D if and only if she is \textit{competent} with respect to that domain, and she would take the fact that A is \textit{counting on her}, were A to do so in this domain, to be a \textit{compelling reason} for acting as counted on.'' \citep[p.~61]{Jones2012-aw}

Jones’ account points towards three features that could possibly be captured in an observable signal: (i) competence in a domain, (ii) knowledge of a person’s reliance, and (iii) whether that knowledge plays an appropriately causal role in reasoning about how to act. (i) and (ii) point us towards looking first for indications that a system is competent to perform a particular function and recognizes our reliance.\footnote{I use the word ``system'' here to indicate that the same basic reasoning can apply to the socio-technical system more broadly, and (eventually) independently functioning AI systems, though for now I assume these to be insufficiently independent to warrant agent-like consideration as such, as I discuss in the next section.} While ``competence'' and ``recognition'' broadly will take some work to operationalize, there will often be clear prerequisites to look for, such as the presence of certain functionalities and data structures, and past performance in similar situations. The key component to attend to here is (iii): figuring out how to define and detect appropriate inclusion of reliance into a \textit{reasoning} process for action.  Here we might look to work in explainable AI \citep{Miller2018-hr,Miller2019-wc} for guidance.

As \citet{Jacovi2020-pp} argue, our goal should be to achieve \textit{calibrated trust}--trust in systems that is tightly connected to their actual trustworthiness. The paper helpfully distinguishes between mechanisms of \textit{intrinsic trust }that are closely tied to underlying trustworthiness, and \textit{extrinsic trust} that is generated more indirectly by features that are in principle separable from actual trustworthiness. As with ``fakeability'' of signals, these tend to exist on a continuum, but most examples of signals of ``trustworthiness'' unfortunately fall more towards the extrinsic end, including expert testimony and performance on specific test sets and data. Arguably, signals which elicit more \textit{intrinsic} trust should be the goal of transparency mechanisms, as I discuss further below.

\subsection{Transparency}

Calls for transparency are a common theme in ethical AI policy recommendations, and can be thought of as requests or demands for signals of particular material facts about the nature of a particular product or its development context. This framework pushes us to ask which features of a company’s product/process, if made transparent, can function as signals of what facts and towards what ends. As discussed above, one direction to look here is to grant the kind of transparency that would allow users to generate \textit{calibrated trust} in the system by accessing signals tightly tied to the \textit{intrinsic trustworthiness }of the system. 

From an information-theoretic perspective, too much information—the naive aim of total transparency—can be less informative than the right sort of information. Companies can (and do) offer up a lot of true information that is nonetheless misleading--a counterintuitive fact which can be explained by signaling considerations (c.f. earlier discussion about implicature). On the other hand, it is worth exploring whether the raw quantity of data offered transparently could be a blunt but useful signal of beneficial intent \citep{Santos2011-he,Skyrms2002-qz}.

It is also worth considering how the desiderata for beneficence-promoting signaling practices interact with demands of companies to protect intellectual property. There may be a useful analog in arms control \citep{Coe2020-xy}. It is also worth exploring connections with \textit{differential privacy} \citep{Dwork2006-bl}, with an eye towards if and how privacy and the right sort of transparency can co-exist. While this will no doubt be a difficult needle to thread, a signaling perspective might help us formulate transparency demands in a more focused manner than mere maximization, which could make the problem more tractable. 

\subsection{Intentionality}
\label{sec:intent}

The question of trustworthiness is related to broader questions about \textit{intent}, both from corporate agencies and eventually AI systems themselves. When we assess a person as trustworthy, or more generally, when we attempt to predict their behavior and develop a strategic response, a major component of our model will often be our assessment of their \textit{intentions}. We understand this both literally in terms of what, mechanically, people intend to do with their bodies, but also more abstractly, in terms of what guiding values and principles we expect will underlie the particulars of their ultimate mechanical choices. The notion of intention is complicated when it comes to collective/corporate agents \citep{List2011-nt}, and more so when it comes to non-human entities. Characterizing and assessing intention in AI is a key conceptual goal for many researchers in the ``AI Alignment'' sphere, since much of the work in this area presupposes something intention-like is or will be the key target of alignment efforts. 

I caution that overreliance on intention can be predictively and strategically problematic even when interacting with humans. Intention only ever manifests as behavior conditionally on context, to the extent that considerations of context often should dominate (or be considered entangled with) considerations of intent for strategic interaction. Once we acknowledge that even humans lack the kind of ``robust'' notion of intention that is often sought, we see that the need to identify and attribute something like intention to artificial agents is less conceptually relevant than we may have thought. 

The signaling perspective, I argue, refocuses us away from the question of intention (or at least, an overly anthropomorphic concept of it) in ways that may prove instructive. When we look for ``signals'' of beneficent disposition rather than evidence of intention, we do not require an agent to be deliberately producing those signals in order to demonstrate some internal agentic state. This makes us less vulnerable to the previously discussed challenges associated with ``credible commitments''. It also paves the way for a clearer path from the current paradigm of prediction and strategy associated with models of corporate entities developing AI technologies, to future models for which the relevant considerations will increasingly be incorporated into the AI systems themselves. We need not concern ourselves with identifying a phase shift from one to the other upon which we are dealing with an AI ``agent'' with its own ``intention'', requiring a wholly new model to understand and respond. 

\section{Discussion}
\label{sec:disc}

I’ll end by drawing out what I think are the three core lessons to take from considering beneficial AI development dynamics through a signaling lens. I start more abstractly, and end more concretely.

\subsection{AI ethics goes beyond direct consequences.}

Theory and practice around ethical AI generally focuses on the ``first-order'' question of which particular actions lead to better or worse outcomes. The signaling lens shifts our attention to the ``second-order'' question of how our actions influence the ethical behavior of other agents.\footnote{For the ethicists among you: this distinction cross-cuts the consequentialism/deontology debate. While deontologists will likely be more amenable to a relational perspective, they can still be susceptible to a ``first-order'' myopia, focused too closely on fulfilling obligations to others and not on encouraging ethical behavior in others. Consequentialism is also compatible with a second-order perspective that takes a more ``holistic'' approach to consequences in terms of the global, dynamical effects of our actions.} 

Every visible action performed by an individual, team, company, or product, has the potential to influence the beliefs, expectations, and motivations of those who perceive it, and this influence can propagate and have profound social implications. When trying to act ``ethically'', these agents need to not only consider the direct effects of their actions, but the reverberating effects of the signals generated by these actions. This places a higher burden of ethical considerations, but also opens up new avenues of positive impacts the organization can aim towards. 

One of the most promising targets for signaling considerations are the various metrics and benchmarks that are used to quantify and demonstrate adherence to ethical principles such as fairness, privacy, and transparency. These metrics are used in different ways by researchers, regulators, users, and the public to assess the desirability and acceptability of AI products. Depending on how they are understood by these actors and inform their behavior, this generates a straightforward ``reward'' for other companies in the space to ``compete'' with respect to performance on these metrics. It is therefore important to attend to which metrics are used and promoted, and how they are communicated.

\subsection{Signals are dynamic and adaptive}

Signals are ``living'' in the sense that they are unconstrained by our hopes and expectations for them, and they change over time in response to their environment and how we use them. Recognizing this opens up new, innovative ways we might use signals to promote pro-social behavior in AI development contexts. Every output produced in the course of developing products and ensuring their beneficence generates lots of observable information that we can use to identify or manufacture signals that might be suitable for our purposes, which will evolve over time. 

Conversely, if we fail to acknowledge the dynamism of signals we risk to be overconfident in our interpretations of signals, and unprepared for when they change out from under us. For example, any of the various ``fairness'' metrics for classifiers only indicate the presence of ``fairness'' in a narrow and contextual sense. Not only should we be more circumspect in which metrics we use and how we interpret them, we need to recognize that depending on how we use them to inform our behavior, i.e. by ``rewarding'' or ``punishing'' those perform well or poorly on them, we inevitably shift their meanings to reflect that new reward structure.

\subsection{Content arises from contextual incentives}

The central lesson of signaling theory is that the meaning of a signal is entirely captured by the beliefs and motivations of the sender and receiver. In other words, all there is to know about what someone is communicating is who is saying it and why. And to know what is heard, you need to know who is listening, and what they believe about the speaker’s beliefs, intentions and motivations. 

This is a lesson I see my university colleagues grappling with in light of ChatGPT. In the past (to an extent), when a student handed in an essay, they could (or at least did) take the essays at ``face value'' as constituting an honest report of what the student has learned and demonstrating their best attempt to respond to the prompt. Knowing that ``the same'' content could have been generated merely by copying and pasting an essay prompt into a browser has completely undermined their faith in that signal. 

The problem professors are contending with is that they \textit{thought} they were asking students to write an essay in response to their prompt, but what they \textit{actually} communicated to students was to generate some unique artifact that meets certain criteria indicated on the syllabus—a task they can accomplish without engaging critically with the course content. The effect of ChatGPT here was to widen the gap between the assumed association between ``essay with certain characteristic'' and ``student who learned the material'' sufficiently to break the signaling association. ``Solving'' this problem thus requires mending the gap, either by (i) enforcing compliance through surveillance or auditing, (ii) designing the assignment so that doing well, despite the ability to use ChatGPT, still requires students to ``incidentally'' learn, or (iii) shifting students incentives to be more intrinsically associated with learning than doing well on assignments. 

As an educator, I strive for (iii), acknowledge that (ii) is often more realistic though quite challenging, and view (i) as a dangerous stopgap that positions students as adversaries, which I hope to avoid whenever possible. Achieving (iii) requires being open and curious about what students' actual interests and goals are, and working flexibly with them to find ways to find learning objectives that resonate with them. The approach for corporate actors will of course be different but the principles are the same. If we rely solely on adversarial, punitive measures (i), we essentially ensure an eternal conflict in which companies inevitably exploit and enhance loopholes until our methods are rendered useless. Pure alignment of values (iii) is a worthy aspiration but basically impossible to work towards directly. Our best bet is to focus on carefully creating and adjusting demands and expectations so that the most ``rational'' response to incentives is to behave prosocially.

\bibliographystyle{chicago}
\bibliography{bibliography-bibtex}

\end{document}